\documentclass[prb,superscriptaddress,letter,10pt,twocolumn]{revtex4}
\usepackage{amsmath}
\usepackage{amssymb}
\usepackage{graphicx}
\usepackage{color}
\usepackage{epsfig}
\usepackage{dcolumn}
\usepackage{bm}
\usepackage{bm}
\usepackage{blindtext}
\usepackage{natbib}
\usepackage{epstopdf}
\usepackage{ulem}
\usepackage{url}
\usepackage{bm}
\usepackage{subfigure}

\def\be{\begin{equation}}
\def\ee{\end{equation}}
\def\bea{\begin{eqnarray}}
\def\eea{\end{eqnarray}}

\def\ba{\begin{aligned}}
\def\ea{\end{aligned}}

\def\arccosh{\text{arccosh}}

\newcommand\w           {\omega}

\newcommand\p             {\partial}

\newcommand{\vt}{\vartheta}

\newcommand{\wx}[1]{{\color{black} #1}}

\newcommand{\mk}[1]{{\color{black} #1}}

\begin{document}
\title{Cascade of singularities in the spin dynamics of a perturbed quantum critical Ising chain}
\author{Xiao Wang}
\affiliation{Tsung-Dao Lee Institute and School of Physics \& Astronomy, Shanghai Jiao Tong University, Shanghai, 200240, China}
\author{Haiyuan Zou}
\affiliation{Tsung-Dao Lee Institute and School of Physics \&  Astronomy, Shanghai Jiao Tong University, Shanghai, 200240, China}
\author{Krist\'{o}f H\'{o}ds\'{a}gi}
\affiliation{BME ``Momentum" Statistical Field Theory Research Group, Institute of Physics, Budapest University of Technology and Economics, 1111 Budapest, Budafoki \'{u}t 8, Hungary}
\author{M\'{a}rton Kormos}
\affiliation{MTA-BME Quantum-Dynamics and Correlations Research Group,  E\"otv\"os Lor\'and Research Network (ELKH), Budapest University of Technology and Economics, 1111 Budapest, Budafoki \'{u}t 8, Hungary}
\author{G\'abor Tak\'acs}
\affiliation{MTA-BME Quantum-Dynamics and Correlations Research Group,  E\"otv\"os Lor\'and Research Network (ELKH), Budapest University of Technology and Economics, 1111 Budapest, Budafoki \'{u}t 8, Hungary}
\author{Jianda Wu \footnote{Electronic address: wujd@sjtu.edu.cn}}
\affiliation{Tsung-Dao Lee Institute and School of Physics \& Astronomy, Shanghai Jiao Tong University, Shanghai, 200240, China}

\begin{abstract}
When the quantum critical transverse-field Ising chain is perturbed by a longitudinal field, a quantum integrable model emerges in the scaling limit with massive excitations described by the exceptional $E_{8}$ Lie algebra. Using the corresponding analytical form factors of the quantum $E_{8}$ integrable model, we systematically study the spin dynamic structure factor of the perturbed quantum critical Ising chain, where particle channels with total energy up to 5$m_1$ ($m_1$ being the mass of the lightest $E_{8}$ particle) are exhausted. In addition to the significant single-particle contributions to the dynamic spectrum,
each two-particle channel with different masses
is found to exhibit an edge singularity
at the threshold of the total mass
and decays with an inverse square root of energy, which is attributed to the singularity of the two-particle density of states at the threshold.
The singularity is absent for particles with equal masses due to a cancellation mechanism involving the structure of the form factors.
As a consequence, the dynamic structure factor
displays a cascade of bumping peaks in the continuum region with clear singular features which can serve as a solid guidance for the material realization of the quantum $E_{8}$ model.
\end{abstract}
\maketitle

\section{introduction}
Due to collective quantum fluctuations,
exotic states of matter can emerge
near a quantum critical point (QCP)
in quantum many-body systems~[\onlinecite{sachdev_2011,BELAVIN1984333, 1996tyli.conf..559Z, 2004NuPhB.676..587B,DELFINO199440,PhysRevD.19.2477}],
which have been attracting
intensive research~[\onlinecite{JLTP_2010,Si_Nature_2001,
Coleman2005,Schuberth485,PhysRevLett.113.247201,Wang_2018,PhysRevB.97.245127,PhysRevLett.123.067202,PhysRevLett.123.067203,PhysRevResearch.2.013345}].
One such paradigmatic system is the transverse field Ising chain (TFIC)
in the presence of a longitudinal field.
At the QCP of the TFIC, conformal invariance emerges in the scaling limit,
corresponding to a central charge 1/2 conformal field theory (CFT). Turning on
a small longitudinal field at the QCP gives a perturbation
to the conformal field theory, resulting in
a massive relativistic field theory model
with an emergence of eight stable particles of masses
$m_i, i=1, \cdots, 8$ (Fig.~\ref{Fig.1})~[\onlinecite{zam}]. The mass ratios
of the eight particles and
their scatterings and form factors
are beautifully organized
by the $E_8$ exceptional Lie algebra, dubbed as
the quantum $E_8$ integrable model ~[\onlinecite{inbook,zam}].

A material realization has been long sought since
the discovery of the quantum $E_8$ integrable model. One decade ago,
inelastic neutron scattering measurements
in quasi-one-dimensional (1D) ferromagnetic CoNb$_2$O$_6$
provided
preliminary evidence for the
lowest two states of the
quantum $E_8$ spectrum  corresponding to the lightest two particles with masses $m_1$ and $m_2$~[\onlinecite{2018Quantum}],
which further motivated material-based studies of this
exotic system~[\onlinecite{PhysRevB.83.020407}]. Recently, a combination of theoretical
and experimental efforts led to a full realization of the
quantum $E_8$ spectrum in the material of
BaCo$_2$V$_2$O$_8$ (BCVO)~[\onlinecite{PhysRevB.101.220411,Zou:2020ouw}].
In Refs.~[\onlinecite{PhysRevB.101.220411,Zou:2020ouw}],
besides a direct physical instruction to concretely guide
the experimental realization
of the quantum  $E_8$ spectrum in BCVO,
we also provided a summary of an analytical form factor framework
to determine the corresponding dynamical structure factor (DSF).
The analytical DSF data in Refs.~[\onlinecite{PhysRevB.101.220411,Zou:2020ouw}]
have been broadened in accord with realistic experimental energy resolution.
The excellent agreement implies the first experimental realization of
the quantum $E_8$ integrable model in a real material.
Motivated by this exciting progress,
in this paper we give a complete account of the details of the analytic calculations, greatly expanding the discussion presented in the recent experimental-theoretical work [\onlinecite{PhysRevB.101.220411,Zou:2020ouw}] on the material realization of the quantum $E_{8}$ model in BaCo$_2$V$_2$O$_8$.


When unfolding details of the analytical framework, we uncover rich features
not revealed in Refs.~[\onlinecite{PhysRevB.101.220411,Zou:2020ouw}], such as
the singular structure of the DSF spectrum
smeared in the
broadened analytical DSF data~[\onlinecite{PhysRevB.101.220411,Zou:2020ouw}].
We find that besides the well-known singularities
from the single-$E_8$-particle channels, the two-particle channels with different
masses lead to a cascade of edge singularities in the dynamic spectrum, where
the threshold for
each edge singularity is the total mass of the two particles. When
energy \
is beyond the edge-singularity threshold,
the two-particle spectrum decays in a power of an inverse square root.
This singularity can be traced to the divergence
of the two-particle density of states (DOS)
at the threshold which, however, is accidentally
canceled for two-particle channels with equal masses due to the special analytical structure of the form factors. We further demonstrate the smoothness and quick
decrease of the contributions of three and more particle channels to the DSF.
Due to the rapidly decreasing spectral weight and exponentially increasing
CPU cost of carrying out the multi-fold integration
with increasing particle number,
we choose the energy cutoff at 5$m_1$ and focus on zero total momentum
in all DSF calculations. The DSF for non-zero momentum as
well as the corresponding dispersion are deferred to
future study.

The rest of the paper is organized as follows.
Sec.~II elaborates the necessary ingredients of our calculations.
Sec.~III provides analytical calculations in detail.
Then we discuss our results with experimental realization
and draw conclusion in Sec.~IV.
The details of the analytic calculations can be found in the Appendices,
including
the complete and correct set of recursive equations
for systematically obtaining the form factors of the quantum $E_8$
model.

\section{The model}
\label{sec2}

\begin{figure}[t!]
	\centering
		\includegraphics[width=8cm]{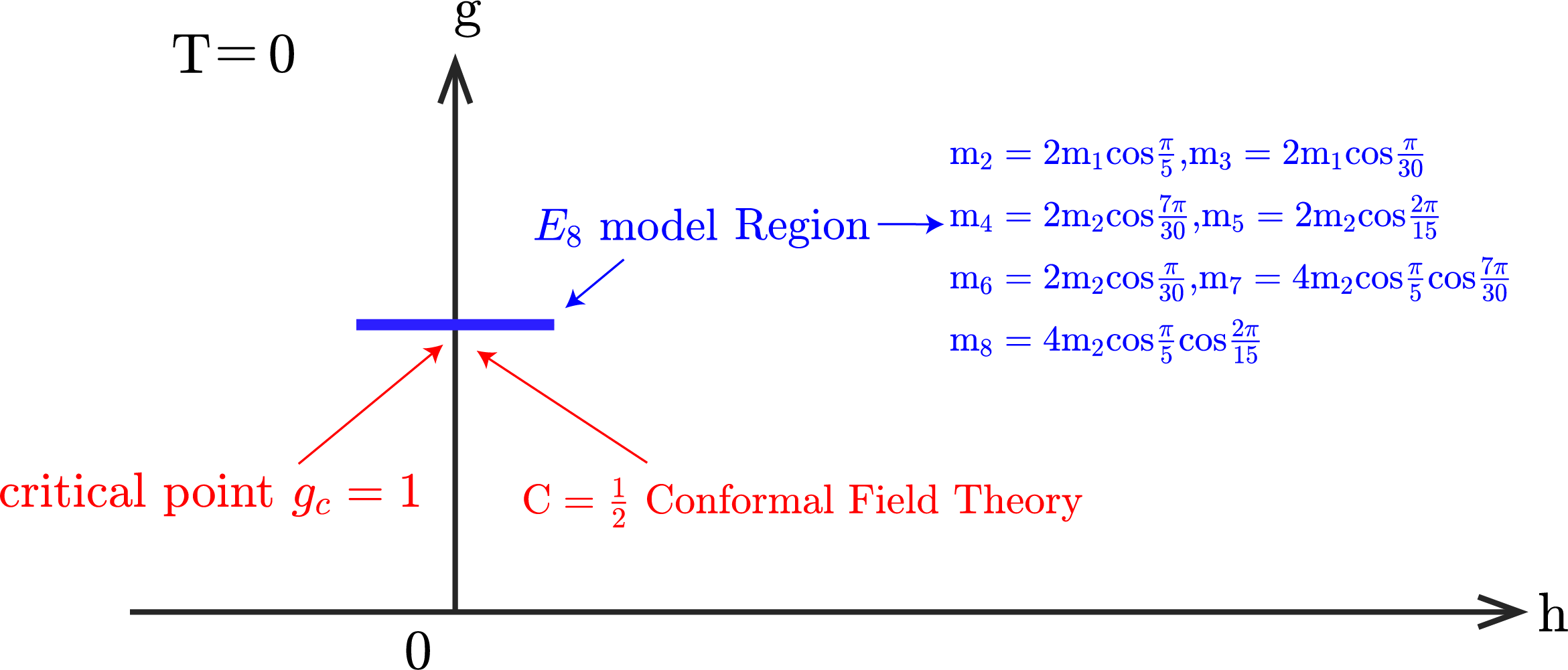}\\
	\centering
	\caption{The blue solid line illustrates the region $E_8$ physics emerges, where
	parameters $h$ and $g$ for horizontal and vertical axes
	are from the Hamiltonian Eq.~(\ref{2}). The masses of particles
	are displayed in terms of the lightest two masses $m_1$ and $m_{2}.$}
	\label{Fig.1}
\end{figure}

We consider the transverse-field Ising chain (TFIC) at its QCP $g=g_c=1$
perturbed by a longitudinal field $h_{z}$,
\be
H_{\text{pert}}=-J(\sum_{i}\sigma_{i}^{z}\sigma_{i+1}^{z}+\sum_{i}\sigma_{i}^{x}+h_{z}\sum_{i}\sigma_{i}^{z}),
\label{2}
\ee
where $J > 0$, and $\sigma_i^\alpha$ at site $i$ are the
Pauli matrices related to the spin operators
$S_i^{\alpha}=\sigma_i^{\alpha}/2,$ $(\alpha=x,y,z)$ with the
Planck constant set to $ \hbar=1$. \mk{In the scaling limit, the QCP is described by the conformal field theory of central charge $c=1/2$ corresponding to free massless fermions. The Hamiltonian \eqref{2}
gives rise to a field theory
obtained by perturbing the
CFT by one of its relevant primary
operator~[\onlinecite{inbook,PhysRevLett.113.247201,PhysRevD.18.1259}]},
\be
H_{E_{8}}=H_{c=1/2} - h\int d x\,\sigma(x)\,,
\label{HE8}
\ee
\mk{where the operator $\sigma(x)$ and the field $h$ are the rescaled field theory versions of the lattice magnetization operator $\sigma_i^z$ and longitudinal magnetic field $h_z,$ respectively (for the precise relations c.f. Appendix \ref{app:scaling}).

As shown by Zamolodchikov~[\onlinecite{zam}],
this perturbation opens a gap and leads to an integrable quantum field theory, the so-called $E_{8}$ model. The hallmark of this model
is the presence of
eight stable particle excitations~[\onlinecite{DELFINO1995724}]
with the mas of the lightest particle is $m_{1}\sim\vert h \vert^{8/15}$
and all the other masses can also be expressed in terms of $m_1$ exactly, as shown in Fig.~\ref{Fig.1}.}

\begin{figure}[t]
	\centering
	\includegraphics[width=8cm]{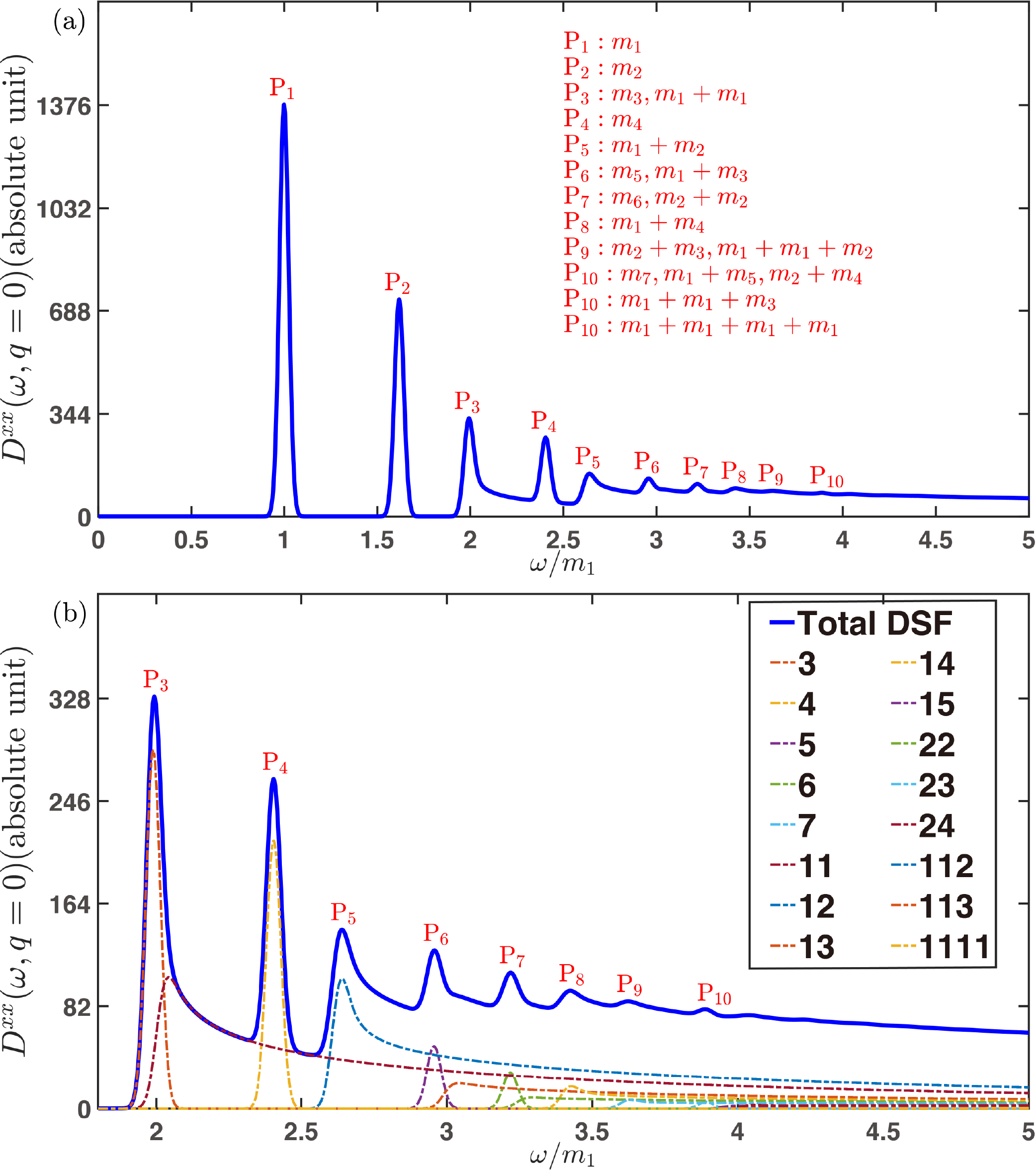}
	\centering
	\caption{The total $D^{xx}(\w,q=0)$
intensity as a function of $\w$ with 0.05$m_{1}$ broadening.
(a) Several peaks associated with single or multi-particle excitations are labelled by $P_i$,
where $i=1$ to 10 as $\w$ increases.
(b) The details of spectrum with excitation energy larger than $2 m_1$. Each dashed curve is the DSF from the corresponding channel.
The numbers in the legend are abbreviated labels for $E_8$ particles. For example, ``11" stands for $m_1 + m_1$ channel. }
	\label{Fig.2}
\end{figure}
The energy and momentum eigenstates of the Hamiltonian can be written as asymptotic states $\vert A(\theta_{1}),...,A(\theta_{n})\rangle_{a_{1},...,a_{n}}$ with the orthogonality and normalization relations
\be
_{a_{i}}\langle A(\theta_{i})\vert A(\theta_{j})\rangle_{a_{j}}=2\pi \delta_{a_{i}a_{j}}\delta (\theta_{i}-\theta_{j}),
\ee
where $\vert A(\theta_{i})\rangle_{a_{i}}$ ($a_{i}=1,\cdots, 8$)
labels a state of an $E_8$ particle with mass $m_{a_{i}}$ and
rapidity $\theta_i$. The energy and momentum eigenvalues
written in terms of the relativistic rapidity parameter $\theta$ are $E=\sum_{i=1}^{n}m_{a_{i}}\cosh(\theta_{i})$ and $P=\sum_{i=1}^{n}m_{a_{i}}\sinh(\theta_{i})$, respectively.

In the following, we consider the two point correlation function of the operators $\Phi = \sigma^{x,y,z}$,
\be
\langle \Phi(x,t)\Phi(0,0)\rangle =\langle 0\vert e^{-iPx}e^{iHt}\Phi(0,0)e^{-iHt}e^{iPx}\Phi(0,0)\vert 0\rangle
\ee
where $\vert 0\rangle$ stands for the ground state (vacuum)
of the $E_{8}$ Hamiltonian.
By inserting a complete basis of the $E_{8}$ eigenstates
into the correlation function, the dynamic structure factor (DSF) with
zero momentum transfer expressed in the Lehmann representation follows as
\begin{multline}
\label{new2}
D^{\Phi\Phi}(\omega,q=0)\\
=\sum_{n=0}^\infty\sum_{\{a_i\}}(\prod_{a_{i}}\dfrac{1}{N_{a_{i}}!})\dfrac{1}{(2\pi)^{n-2}}  \int_{-\infty}^{\infty} d\theta_{1}...d\theta_{n}\\
|\langle0|\Phi\vert A(\theta_{1}),...,A(\theta_{n})\rangle_{a_{1},...,a_{n}}|^2
\delta(\omega-\sum_{i=1}^{n}E_{i})\delta(\sum_{i=1}^{n}p_{i}).
\end{multline}
As Eq.~(\ref{new2}) shows, the dynamic properties of the system are
determined by the combined effects from the on-shell particles with
total energy and momentum conservation.
By choosing different number of particles
in the complete basis, the contributions of the DSF
can be divided into different channels: single-, two-, three-particle channels
and so on. Each channel's DSF contribution exhibits
special features as shall be discussed in the following sections.
To calculate Eq.~(\ref{new2}), the form factors
\be
F_{a_1,...,a_n}^{\Phi}(\theta_1,...,\theta_n)=\langle 0|\Phi| A(\theta_{1}),...,A(\theta_{n})\rangle_{a_{1},..,a_{n}},
\label{new3}
\ee
are needed, which can be derived following the form factor bootstrap
approach~[\onlinecite{DELFINO1995724,DELFINO1996450,H_ds_gi_2019,
Schuricht_2012,Bertini_2014,Cort_s_Cubero_2017}].
The detailed form of the recursive equations and a discussion of the method
of solving them are presented in Appendix \ref{app:FFrecursive}.

\mk{For practical calculations, the infinite form factor series must be truncated.} In this work, we systematically calculate
the form factor contributions up to the energy cutoff at $5m_{1}$.


\section{The spin dynamic structure factor of the quantum $E_8$ integrable model}
We now proceed to calculate the
DSF $D^{\Phi\Phi}(\omega, q=0)$ with $\Phi=\sigma^i$ and $i=x,y,z$
(abbreviated as $D^{ii}$) for the $E_8$ model.
There are two relevant \mk{operators, $\sigma (x)$ (magnetization density)}
and $\epsilon (x)$ (energy density)
in the $E_8$ model, corresponding to $\sigma_i^z$ and $\sigma_i^x$
in the lattice model~[\onlinecite{DELFINO1995724}], respectively \mk{(see also Appendix \ref{app:scaling})}. As a result,
within the framework
of quantum $E_8$ integrable model, one is only able to
determine $D^{xx}$ and $D^{zz}$.
$D^{yy}$ can be determined through an exact relation
$D^{yy}(\w)=\w^{2}D^{zz}(\w)/(4J^{2})$~[\onlinecite{PhysRevLett.113.247201}].
$D^{xx}(\omega, q=0)$ is shown in Fig.~\ref{Fig.2},
and the results for $D^{yy}(\omega, q=0)$ and $D^{zz}(\omega, q=0)$
can be found in the Appendix \ref{app:DSFofYYZZ}.~[\onlinecite{SM}].

For illustration, in Fig.~\ref{Fig.2} we broaden the DSF
with an energy resolution
of $0.05 m_1$.
When the transferred energy is larger than $2 m_1$,
multi-particle excitations appear.
Remarkably, the high energy excitations
retain visibility in the DSF continuum region (Fig.~\ref{Fig.2})
up to $m_7$.
In particular, the various two-particle spectrum contributions leave
significant ``resonant" features
with bumpy peaks
in the continuum region of the spectrum,
whose origin will be discussed in detail
in Sec.~\ref{subsectwo}. The clear observation
of this theoretically expected series of two-particle peaks
at the corresponding transferred energy
in the material of BCVO~[\onlinecite{PhysRevB.101.220411,Zou:2020ouw}]
provides a smoking gun signature for the material realization of the $E_8$ model.

In the following, to analyze contributions from the
single and multi-particle excitations in detail,
we specify the contributions of the different channels according to
the number of particles and exhaust all possible cases with
energy less than $5m_1 > m_8$.

\subsection{Single-particle channels}
\begin{figure}[t]
	\centering
	\includegraphics[width=8cm]{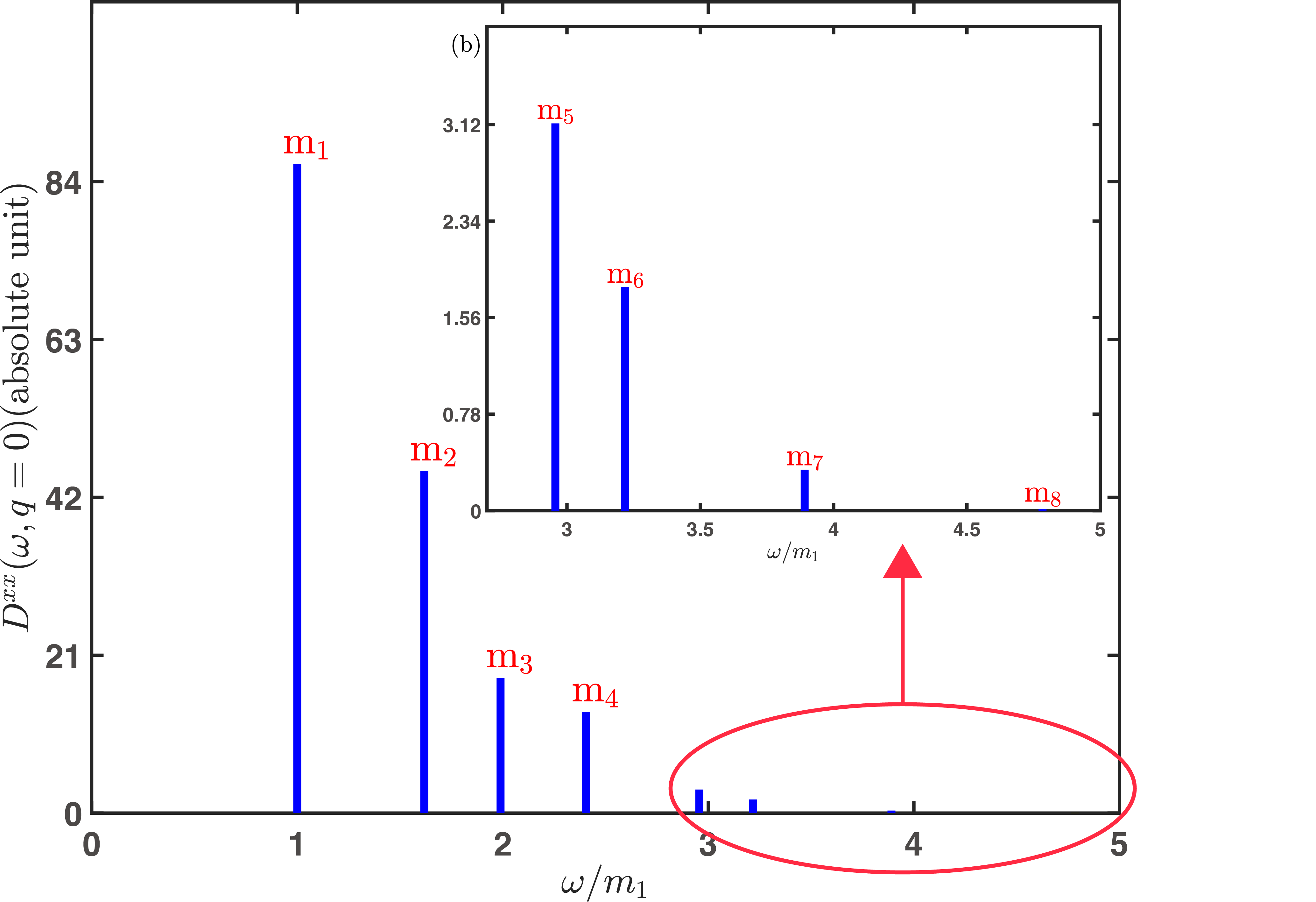}
	\centering
	\caption{$D^{xx}_1(\omega, q=0)$ contribution from the single particle
	channels are delta functions with different spectral weights for different $E_8$ particle species.
	The inset exhibits the spectral weights from $m_{5}$ through $m_{8}$.}
	\label{Fig.3}
\end{figure}
The single-particle contributions to DSF are given by
\be
D^{\Phi\Phi}_1(\omega,q=0)=\sum_{i=1}^{8}2\pi\dfrac{\vert F^{\Phi}_{a_{i}}\vert^{2}}{m_{a_{i}}}\delta(\w-m_{a_{i}}),
\label{11}
\ee
where the form factor $F_{a_i}^\Phi$ is the single particle
form factor for the $E_8$ particle $a_i$~[\onlinecite{DELFINO1996450}],
and $2\pi \vert F^{\Phi}_{a_{i}}\vert^{2}/{m_{a_{i}}}$ gives the corresponding single-particle spectral weight.
In Fig.~\ref{Fig.3} we show the spectral weight
of $D^{xx}_1$ for each $E_8$ particle.

\subsection{Two-particle channels} \label{subsectwo}
From Eq.~(\ref{new2}), we get following two-particle contributions
to the DSF,
\be
\ba
D^{\Phi\Phi}_2(\omega,q=0)=\sum_{i\leq j}(\dfrac{1}{2})^{\delta_{a_{i}a_{j}}}\dfrac{\vert F^{\Phi}_{a_{i}a_{j}}(\theta_{1}-\theta_{2}) \vert^{2}}{m_{a_{i}}m_{a_{j}}\vert \sinh(\theta_{1}-\theta_{2}) \vert},
\label{dsf2}
\ea
\ee
where
\be
\theta_{1}-\theta_{2}=\arccosh\left(\dfrac{\omega^2-m_{a_{i}}^2-m_{a_{j}}^2}{2m_{a_{i}}m_{a_{j}}}\right)\,,
\label{13}
\ee
with the lower bound of energy as $\omega_{min}=m_{a_i}+m_{a_j}$,
the spectrum threshold for
a specific two-particle channel with masses $m_{a_i}$ and $m_{a_j}$.

Fig.~4 shows the analytical two-particle DSF results
of $D^{xx}_2$ by considering all possible combinations with $\omega<5m_1$.
Edge singularities are exhibited for two-particle channels with different masses.
In Eq.~\eqref{dsf2}, the
Jacobian term $1/\vert\sinh(\theta_1-\theta_2)\vert,$
which is just the two-particle density of states,
contributes a singular behavior at $\theta_1=\theta_2$,
corresponding to $\omega=\omega_0$ ($\omega_0 = m_{a_i}+m_{a_j}$).
A simple analysis gives $D_{2}^{xx}\sim 1/\sqrt{\omega-\omega_0}$ at $\omega\gtrsim \omega_0$
~[\onlinecite{SM}]. Apparently, this singular behavior also appears
in $D^{yy}$ and $D^{zz}$.
This scaling behavior is further illustrated in Fig.~4(b)
by the logarithmic fitting of $D^{xx}$ at $m_1m_2$ channel
with fitting $(\omega-\omega_0)^\alpha$, $\alpha=-0.5043$.
The edge singularity disappears for two particles with the same mass,
which is due to the explicit form of equal-mass-two-particle form factor where a
$|\sinh^2(\theta_1-\theta_2)/2|$ term appears and cancels
the $\sinh(\theta_1-\theta_2)$ in the denominator of Eq.~(\ref{dsf2}) [\onlinecite{SM}].
\begin{figure}
	\centering
	\includegraphics[width=7.3cm]{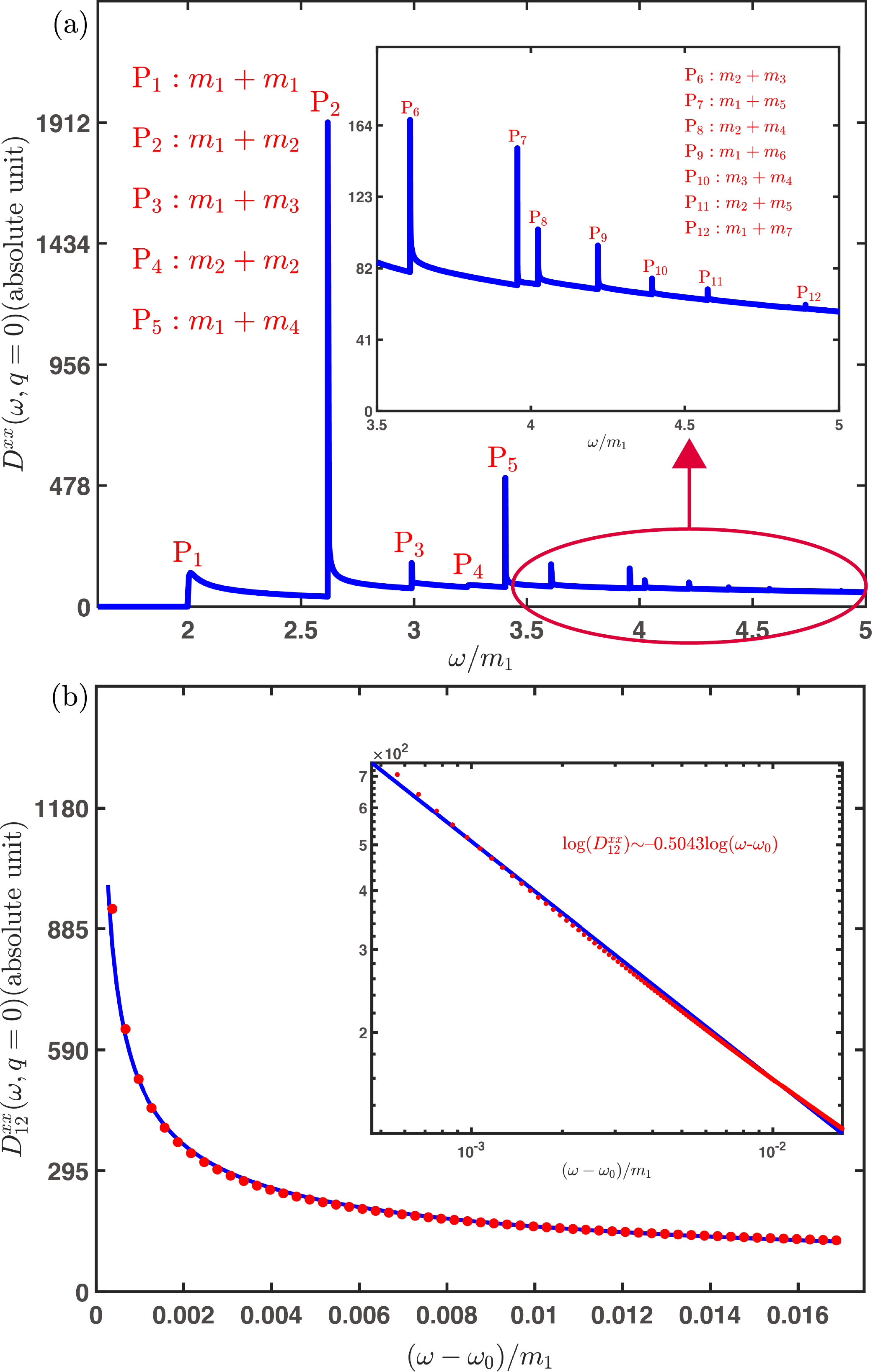}
	\centering
	\caption{The two-particle DSF. (a) $D^{xx}_2(\omega, q=0)$ contributed from all
	two-particle channels. The  $\omega>3.5m_1$ region is highlighted in the inset.
	(b) The singularity near the peak arise at $\w_{0}=m_{1}+m_{2}$ has an inverse square root $\sim (\w-\w_{0})^{-1/2}$ behavior, confirmed by the logarithmic fit for the $m_1+m_2$ channel shown in the inset.}
	\label{Fig.4}
\end{figure}

\subsection{Three and four-particle channels}
\begin{figure}
	\centering
	\includegraphics[width=7.15cm]{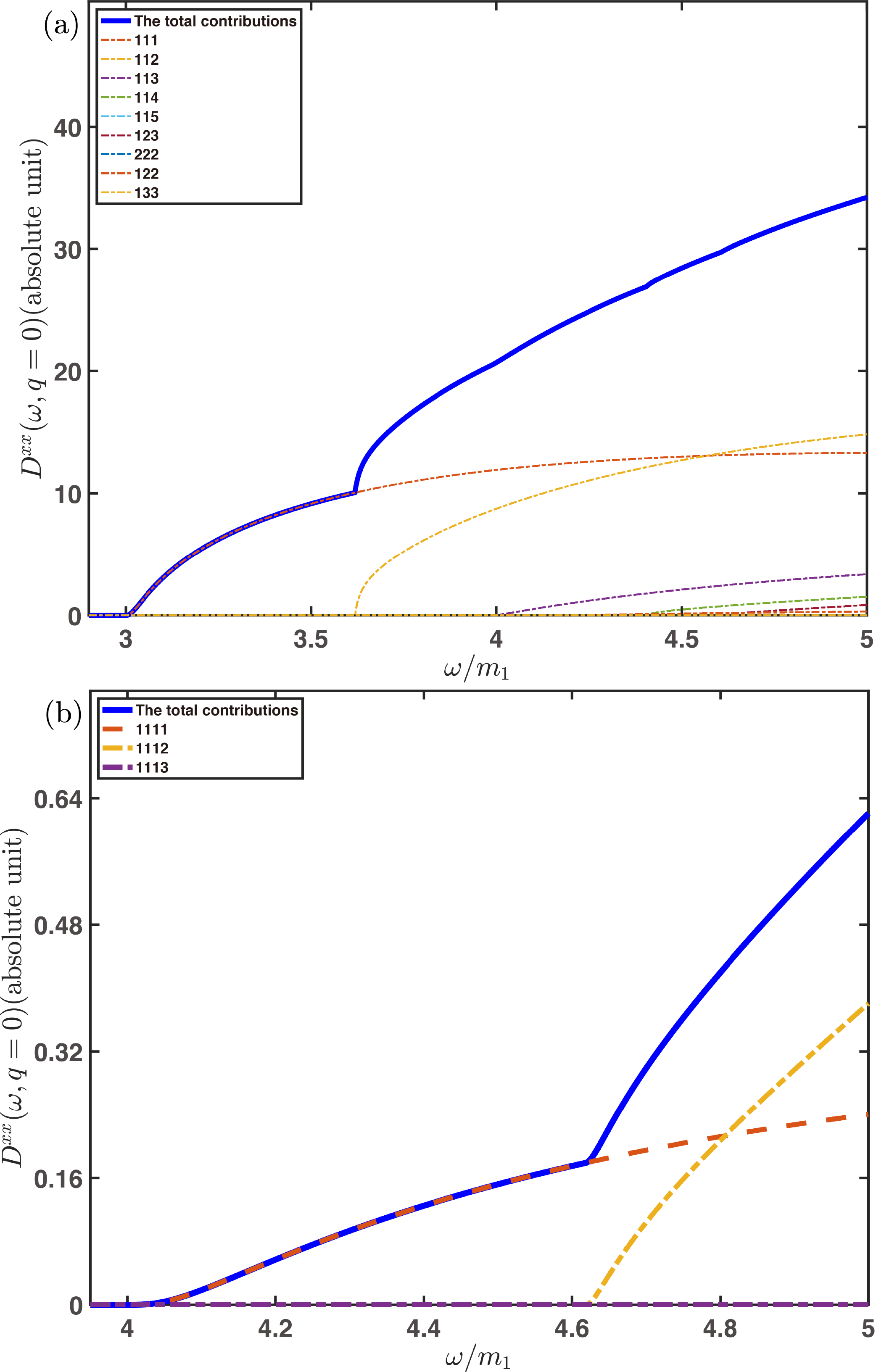}
	\centering
	\caption{$D^{xx}(\omega, q=0)$ from three-particle channels (a) and four-particle channels (b) for $\omega<5m_1$.}
	\label{Fig.5}
\end{figure}
In this section we further determine the contributions
of multi-particle channels beyond two particles.
From Eq.~(\ref{new2}), we have DSF contributions
\be
\ba
D^{\Phi\Phi}_3(\omega,q=0)&=\sum_{i\leq j \leq k}(\prod_{a_{i,j,k}}\dfrac{1}{N_{a}!})\dfrac{1}{(2\pi)}\\
&\times\int_{-\infty}^{\infty} d\theta_{3}\dfrac{\vert F^{\Phi}_{a_{i}a_{j}a_{k}}(\theta_{1},\theta_{2},\theta_{3}) \vert^{2}}{m_{a_{i}}m_{a_{j}}\vert \sinh(\theta_{1}-\theta_{2}) \vert}
\label{14}
\ea
\ee
for the three-particle channels, and
\be
\ba
&D^{\Phi\Phi}_4(\omega,q=0)=\sum_{i\leq j \leq m \leq n}(\prod_{a}\dfrac{1}{N_{a}!})\dfrac{1}{(2\pi)^2}\\
&\times \int_{-\infty}^{\infty} d\theta_{3}d\theta_{4}\dfrac{\vert F^{\Phi}_{a_{i}a_{j}a_{m}a_{n}}(\theta_{1},\theta_{2},\theta_{3},\theta_{4}) \vert^{2}}{m_{a_{i}}m_{a_{j}}\vert \sinh(\theta_{1}-\theta_{2}) \vert}
\label{15}
\ea
\ee
for the four-particle channels, whose results are shown in Fig~\ref{Fig.5}(a,b).
Note that in Eqs. (10)-(11) $\theta_1-\theta_2$ is not the same as in Eq. (9)
but is a function of the integration variables and $\w$.

The threshold of the spectrum for each channel is the total mass, and there is
no singularity~[\onlinecite{SM}]. Compared with the single and two-particle channels,
the spectral weight is two to three orders smaller, which only
slightly modifies the total DSF spectrum shown in Fig.~2.

\section{Discussion and conclusion}
\label{sec4}
In this article, we provided a systematic theoretical analysis
that greatly expands the
theoretical treatment of Refs.~[\onlinecite{PhysRevB.101.220411,Zou:2020ouw,PhysRevB.102.104431}]
and which will also be helpful in guiding a
realization of the $E_8$ model in other possible materials,
such as CoNb$_2$O$_6$~[\onlinecite{PhysRevB.102.104431}].
In BCVO, the QCP of TFIC universality is hidden in the 3D
ordered phase with an inter-chain interaction serving as the
longitudinal perturbation\wx{~[\onlinecite{Zou:2020ouw}]}. The obtained theoretical DSF
can be directly measured by terahertz spectroscopy
measurements~[\onlinecite{PhysRevB.101.220411,PhysRevB.102.104431}]
as well as inelastic neutron scattering experiments~[\onlinecite{Zou:2020ouw}].
The experimentally measured differential cross section
is related to the DSF by~[\onlinecite{chaikin_lubensky_1995}]
\be
\frac{d^2\sigma}{d\Omega dE}\sim\frac{|q_f|}{|q_i|}\sum_{\mu,\nu=x,y,z}(\delta_{\mu\nu}-\frac{Q_\mu Q_\nu}{|\mathbf{Q}|^2})D^{\mu\nu}(\omega,\mathbf{Q}),
\ee
where the scattering vector $\mathbf{Q}$ is defined
as $\mathbf{Q}=q_f-q_i$ with $q_i$ and $q_f$ as initial and final wave vectors,
respectively.
For zero transferred momentum $q_f=q_i + G$
with $G$ being the reciprocal lattice vector of the crystal. After multiplying by a factor to convert the field theory results to
the lattice system,
\be
\langle \sigma_{i}^{\alpha}(t)\sigma_{0}^{\alpha}(0)\rangle_{\text{lattice}} = \langle \sigma^{\alpha}(x,t)\sigma^{\alpha}(0,0)\rangle \cdot (\langle 0 \vert \sigma^{\alpha} \vert 0 \rangle_{\text{lattice}})^{2},
\ee
where $\alpha = x,z$, our DSF results can be compared with a zone center inelastic neutron scattering
or other spectroscopy experiment on $D^{xx}$, $D^{yy}$, and $D^{zz}$.

To conclude, by using the exact analytic form factors of the quantum $E_8$ integrable model,
we performed a detailed calculation of
the DSF of the model with zero total momentum and total transferred energy up to $5m_1$.
The obtained DSF describes the spin dynamics of the TFIC at its QCP with a longitudinal magnetic field perturbation.

In addition to the eight single-particle resonant peaks,
the two-particle DSF contributions with different masses exhibit edge singularities
at the thresholds and decay with an inverse square root behavior.
For the channels involving more than two particles, there is no
such singularity and their spectral contribution decreases quickly
with increasing energy and particle numbers.
The obtained DSF
displays rich and fine spectrum structure with a series of peaks not only from
single particles but also two particles, and especially two-unequal-mass
particles. Thus the obtained DSF fine structure for the $E_8$ model
can guide and evince the material realization of the model
BCVO~[\onlinecite{PhysRevB.101.220411,Zou:2020ouw}].
In the future, based on the current theoretical results and
calculation techniques,
we plan to study the $E_8$ DSF with finite momentum to extract the
dispersion relation and explore physics beyond integrability.

\section*{Acknowledgments}
We thank G. Mussardo for helpful discussions.
The work at Shanghai Jiao Tong University is sponsored by
Natural Science Foundation of Shanghai with Grant No.
20ZR1428400 and Shanghai Pujiang Program with Grant No.
20PJ1408100 (X.W. H.Z. J.W) and the National Natural Science Foundation of China Grant No. 11804221 (H.Z.).
J.W. acknowledges additional support from a Shanghai talent
program.
This work was partially supported by National Research, Development and
Innovation Office (NKFIH) under the research grant K-16 No. 119204 and
also within the Hungarian Quantum Technology National Excellence
Program, project no. 2017-1.2.1- NKP- 2017-00001, and by the Fund
TKP2020 IES (Grant No. BME-IE-NAT), under the auspices of the Ministry
for Innovation and Technology. M.K. acknowledges support by a Bolyai J\'anos grant of the HAS. K.H. was supported by the \'UNKP-20-3, while M.K. was supported by the \'UNKP-20-5 new National Excellence Program of the Ministry for Innovation and Technology from the source of the National Research, Development and Innovation Fund.

\bibliography{main}
\bibliographystyle{apsrev-nourl}


\appendix

\onecolumngrid

\section{Scaling limit of the Ising spin chain}
\label{app:scaling}

In the scaling limit, the lattice constant $a$ is sent to zero with the coupling $J$ sent to infinity and $h_z$ sent to 0 in the manner
\begin{align}
a&\to 0\,,\qquad J\to\infty, \qquad h_z\to0\,,\\
\Delta&\sim J h_z^{8/15}=\text{fixed,}\qquad 2Ja=\hbar c=\text{fixed,}
\end{align}
where $\Delta$ is the energy gap and $c$ is the effective speed of light in the field theory.
Using the field theory normalization conventions and $\hbar=c=1,$ the relations between the magnetic fields and oparators read~[\onlinecite{Pfeuty}]
\begin{align}
h& = \frac2{\bar s}J^{15/8} \,h_z\,,\\
\sigma(x=ja) &=\bar s J^{1/8} \sigma^z_j\,,\\
\varepsilon(x=ja) &=- J^{-1} \sigma^x_j
\end{align}
with $\bar  s = 2^{1/12}e^{-1/8}\mathcal{A}^{3/2}$ where $\mathcal{A}=1.2824271291\dots$ is Glaisher's constant.
The mass of the lightest $E_8$ particle is given in terms of the field $h$ as~[\onlinecite{Fateev}].
\be
m_1 = 4.40490858\, h^{8/15}\,.
\ee

\section{The $E_{8}$ Form Factor Theory}
The definition of the form factor in the $E_{8}$ model has been shown in the main paper, here we recall the definition and give a detailed discussion. The form factor is introduced as
\begin{equation}
F_{a_1,...,a_n}^{\Phi}(\theta_1,...,\theta_n)=\langle 0|\Phi| A(\theta_{1}),...,A(\theta_{n})\rangle_{a_{1},..,a_{n}},
\label{s1}
\end{equation}
where $\theta_i, i=1,\cdots,n$ represent the rapidites, and the asymptotic state
with $n$ particles carries energy and momentum as
\be
E_{n}=\sum_{i=1}^{n}m_{i}\cosh \theta_{i},P_{n}=\sum_{i=1}^{n}m_{i}\sinh\theta_{i}.
\label{s2}
\ee
Let's first focus on the two-particle form factor.
Denote two particles in $m_1, \cdots, m_8$ as $a$, $b$,
the two-particle form factor follows
\be
F_{ab}^{\Phi}(\theta)=\dfrac{Q_{ab}^{\Phi}(\theta)}{D_{ab}(\theta)}F_{ab}^{min}(\theta),
\label{s3}
\ee
where $\theta=\theta_{a}-\theta_{b}$, $Q_{ab}^{\Phi}(\theta)$ are
polynomials in $\cosh\theta$ whose explicit form depends on local operator $\Phi$~[\onlinecite{DELFINO1995724}].
In our calculation, we denote  $Q_{ab}^{1/2}(\theta)$ and $Q_{ab}^{1/16}(\theta)$ for \mk{$\langle \varepsilon(x,t)\varepsilon(0,0) \rangle$ and $\langle \sigma(x,t)\sigma(0,0) \rangle$}, respectively.
In addition,
\be
F_{ab}^{min}(\theta)=\left(-i\sinh(\dfrac{\theta}{2})\right)^{\delta_{ab}}\prod_{\alpha}(G_{\alpha}(\theta))^{p_{\alpha}},
\label{s4}
\ee
where
\be
G_{\alpha}(\theta)=\exp\left\{ 2\int_{0}^{\infty} \dfrac{dt}{t} \dfrac{\cosh(\alpha-t/2)}{\cosh(t/2)\sinh(t)}{\sin}^2\dfrac{(i\pi-\theta) t}{2\pi}\right\},
\label{s5}
\ee
and
\be
D_{ab}(\theta)=\prod_{\alpha}(P_{\alpha}(\theta))^{i_{\alpha}}(P_{1-\alpha}(\theta))^{j_{\alpha}},
\label{s6}
\ee
with
\be
\ba
&i_{\alpha}=n+1, j_{\alpha}=n, \text{if }  p_{\alpha}=2n+1 \\
&i_{\alpha}=n, j_{\alpha}=n, \text{if } p_{\alpha}=2n
\label{s7}
\ea
\ee
and
\be
P_{\alpha}(\theta)=\dfrac{\cos\pi \alpha-\cos\theta}{\dfrac{1}{2}{\cos}^{2}{\dfrac{1}{2}\pi\alpha}}.
\label{s8}
\ee
The parameters $\alpha$ and $p_{\alpha}$
are listed in Table 1 of  Ref.~[\onlinecite{DELFINO1995724}].

Then for the single-particle form factors, using the bound state singularities,
we obtain,
\be
F_{s}^{\Phi}=\frac{\text{Res}(F^{\Phi}_{ab}(\theta)\vert_{\theta=i u_{ab}^{c}})}{i\Gamma_{ab}^{c}}
\label{s9}
\ee
with $\Gamma_{ab}^{c}=\sqrt{i\text{Res}(S_{ab}(\theta)\vert_{\theta=i u_{ab}^{c}})}$. $S_{ab}(\theta)$ refers to the S-matrices in $E_{8}$ theory,
\be
S_{ab}(\theta)=\prod_{\alpha}\left[\frac{\tanh \frac{1}{2}(\theta+i\pi\alpha)}{\tanh \frac{1}{2}(\theta-i\pi\alpha)}\right] ^ {p_{\alpha}}.
\label{s10}
\ee
For arbitary number $n$($n \geq 3$) of particles, the form factor follows by
\be
\ba
F^{\Phi}_{a_{1},...,a_{n}}(\theta_{1},...,\theta_{n})=Q^{\Phi}_{a_{1},...,a_{n}}(\theta_{1},...,\theta_{n}) \prod_{i \leq j}\dfrac{F_{a_{i}a_{j}}^{min}(\theta_{i}-\theta_{j})}{(e^{\theta_{i}}+e^{\theta_{j}})^{\delta_{a_{i}a_{j}}}D_{a_{i}a_{j}}(\theta_{i}-\theta_{j})}.
\label{s11}
\ea
\ee
$Q_{a_{1},...,a_{n}}^{\Phi}(\theta)$ are polynomials in $\cosh\theta$.
Details of $Q^{\Phi}(\theta)$ can be found in Ref~[\onlinecite{Q-web,H_ds_gi_2019}]. In the following section we will give a conclusion to show the main process for the form factor bootstrap method.

\section{form factor Recursive equations and their solution}

\label{app:FFrecursive}

\subsection{The Iteration Process}
We use the following Ansatz for the $n$-particle form factor of the lightest particle $m_1$:
\be
\ba
F_{n}^{\phi}(\vt_1,\vt_2,\dots \vt_n)&\equiv F_{\underbrace{\scriptstyle 1\dots1}_n}^{\phi}(\vt_1,\vt_2,\dots \vt_n)\\
&=H_n\frac{\Lambda_n(x_1,\dots ,x_n)}{(\omega_n(x_1,\dots ,x_n))^n} \prod_{i<j}^{n}\frac{F_{11}^{\text{min}}(\vt_i-\vt_j)}{D_{11}(\vt_i-\vt_j)(x_i+x_j)}\,,
\label{FF1}
\ea
\ee
where $x\equiv\exp(\vt)$ and $\omega_n$ denotes the elementary symmetric polynomials generated by
\be
\prod_{k=1}^{n}(x+x_k)=\sum_{j=0}^{n}x^{n-j}\omega_j(x_1,\dots ,x_n)\,,
\label{FF2}
\ee
and $H_n$ is a constant factor. The operator-dependence is carried by $\Lambda_n(x_1,\dots ,x_n)$ that is an $n$-variable symmetric polynomial that can be expressed in terms of the elementary symmetric polynomials $\omega$. $D_{11}$ can be expressed as
\be
D_{11}(\vt)=P_{2/3}(\vt)P_{2/5}(\vt)P_{1/15}(\vt)\,,
\label{FF3}
\ee
The minimal form factor can be written in the form
\be
F_{11}^\text{min}(\vt)=-i\sinh(\vt/2)G_{2/3}(\vt)G_{2/5}(\vt)G_{1/15}(\vt)\,.
\label{FF4}
\ee
The expression of the recurrence relation is:
\be
\frac{\Lambda_{n+2}(x e^{i\pi/3},x e^{-i\pi/3},x_1,\dots ,x_n)}{x^4\prod_{i=1}^{n}(x-e^{-11 i\pi/15}x_j)(x-e^{11 i\pi/15}x_j)(x+x_j)}=(-1)^n\Lambda_{n+1}(x,x_1,\dots ,x_n)\,,
\label{FF5}
\ee
provided the $H_n$ are chosen to satisfy
\be
\ba
\frac{H_{n+2}}{H_{n+1}}=\frac{\Gamma_{11}^1 \sin \left(\frac{2 \pi }{15}\right) \sin \left(\frac{11 \pi }{30}\right) \sin \left(\frac{8 \pi }{15}\right) \sin \left(\frac{3 \pi }{10}\right)}{2\cos^{2}(\pi/3)\cos^{2}(\pi/5)\cos^{2}(\pi/30)G_{11}(2\pi i/3)}\times\\
\times\left[\frac{\sin^{2}(11\pi/30)\gamma}{4\cos^{2}(\pi/3)\cos^{2}(\pi/5)\cos^{2}(\pi/30)}\right]^n\,.
\label{FF6}
\ea
\ee
The kinematic equation is:
\be
(-1)^n \Lambda_{n+2}(-x,x,x_1,\dots ,x_n)=\mathcal{A}_n U(x,x_1,\dots ,x_n)\Lambda_n(x_1,\dots ,x_n)
\label{FF7}
\ee
with
\be
\ba
U(x,x_1,\dots ,x_n)=\frac{1}{2}x^5&\sum_{k_1,k_2,\dots ,k_6=0}^{n}(-1)^{k_1+k_3+k_5} x^{6n-(k_1+\dots +k_6)}\\
&\times\sin(\frac{\pi}{15}(10(k_1-k_2)+6(k_3-k_4)+(k_5-k_6)))\omega_{k_1}\dots \omega_{k_6}\,,
\label{FF8}
\ea
\ee
and
\be
\mathcal{A}_n=\frac{4\gamma\sin ^2\left(\frac{11 \pi }{30}\right) \left(\cos \left(\frac{\pi }{3}\right) \cos \left(\frac{\pi }{5}\right) \cos \left(\frac{\pi }{30}\right)\right)^2 \left(G_{11}\left(\frac{2 \pi  i}{3}\right)\right)^2}{\left(\Gamma_{11}^1 \sin \left(\frac{2 \pi }{15}\right) \sin \left(\frac{11 \pi }{30}\right) \sin \left(\frac{8 \pi }{15}\right) \sin \left(\frac{3 \pi }{10}\right)\right)^2} \left(\frac{\sin \left(\frac{2 \pi }{3}\right) \sin \left(\frac{2 \pi }{5}\right) \sin \left(\frac{\pi }{15}\right)}{8 \sin ^4\left(\frac{11 \pi }{30}\right) G_{11}(0)\gamma^2}\right)^n\,.
\label{FF9}
\ee

\subsection{Solving For The Two Operators}
The $E_8$ field theory has two scaling fields $\sigma(x)$ and $\varepsilon(x)$ with conformal weights $1/16$ and $1/2$ respectively. Both operators have form factors with polynomial structure determined by the recurrence relations Eq.~(\ref{FF5}) and Eq.~(\ref{FF7}). Consequently, the solution of the forementioned equations is ambiguous in the sense that the general solution corresponds to a field $\phi$ that is the linear combination of the two fields:
\be
\phi = \alpha \sigma + \beta \varepsilon\,.
\label{FF10}
\ee
This means that there are two independent initial conditions from which one can start the recurrence. They were obtained first in~[\onlinecite{DELFINO1996327}] where they made use of the clustering property of form factors which provides the non-linear condition necessary to resolve the linear combination. The clustering property reads~[\onlinecite{DELFINO1996327}]
\be
\lim_{\Lambda\rightarrow\infty} F_{r+l}^\phi (\vt_1+\Lambda,\vt_2+\Lambda,...,\vt_r+\Lambda,\vt_{r+1},...,\vt_{r+l})=\frac{1}{\langle\phi\rangle}F_r^\phi(\vt_1,\vt_2,...,\vt_r) F_l^\phi(\vt_1,\vt_2,...,\vt_l)\,.
\label{FF11}
\ee
Once the initial conditions are known, the solutions for the recurrence can be found up to a single coefficient can be found at each level, the remaining coefficient can be fixed using the clustering property.

In fact, the case for $\sigma(x)$ is even simpler, since it is proportional to the trace of the stress-energy tensor ~[\onlinecite{DELFINO1996450}], so the form factor has to contain a factor $P^+ P^-$ with $P^\pm = \sum_{i=1}^{n}p_i^\pm$ and $p^\pm = p^0\pm p^1 = m (\cosh \vt \pm \sinh\vt)$. Consequently when solving for $\sigma(x)$ the Ansatz can be reduced solving only for symmetric polynomials with $\omega_1(x_1,...,x_n)\neq 0$ and $\omega_{n-1}(x_1,...,x_n)\neq 0$. This modification alone is enough to solve for the polynomials of the $\sigma$ operator without utilizing the clustering property. (This can be checked by verifying the identity $P^+P^-=\omega_1 \omega_{n-1}/\omega_n$, the latter factor coming from the $\omega_n$ in the denominator of the Ansatz Eq.~(\ref{FF1}).)

Nevertheless it has to be used when solving for $\varepsilon$. To this end, one has to calculate the asymptotic behavior of the minimal form factors and the bound state pole factor $D_{11}$. They read:
\be
\lim_{\vt\rightarrow\infty}G_\lambda(\vt) = -i c_\lambda \exp(\vt/2)
\label{FF12}
\ee
where $c_\lambda$ is a real constant that can be obtained from the numerical evaluation of Eq.~(\ref{s5}), so
\be
\lim_{\vt\rightarrow\infty}F_{11}^\text{min}(\vt) = -\frac{1}{2} c_{1/15}c_{2/5}c_{2/3}\exp(2\vt)\,.
\label{FF13}
\ee
For the bound state pole factor we have
\be
\lim_{\vt\rightarrow\infty}D_{11}(\vt) = -\frac{1}{8} \frac{1}{2\cos^{2}(\pi/30)2\cos^{2}(\pi/5)2\cos^{2}(\pi/3)}\exp(3\vt)\,.
\label{FF14}
\ee

These can be combined to impose the constraint coming from the clustering property on the symmetric polynomials, in the simplest case "clustering" only a single rapidity (i.e. $r=1$ and $l=n-1$ in the notation of Eq.~(\ref{FF11})).

\section{The Derivation of The Dynamic Structure Factor}
The two point correlation function for a local
operator $\Phi$ can be organized by the Lehmann representation,
\be
\ba
& \langle 0 \vert \Phi(x,t) \Phi(0,0) \vert 0 \rangle =\langle 0 \vert e^{-iPx} e^{iHt} \Phi(0,0) e^{-iHt} e^{iPx}  \Phi(0,0) \vert 0 \rangle \\
&=(\prod_{a_{i}}\dfrac{1}{N_{a_{i}}!})\int_{-\infty}^{\infty}\dfrac{d\theta_{1}...d\theta_{n}}{(2\pi)^n} \langle 0 \vert e^{-iPx} e^{iHt} \Phi(0,0) e^{-iHt} e^{iPx} \vert \theta_{1},...,\theta_{n} \rangle \langle \theta_{1},...,\theta_{n} \vert \Phi(0,0) \vert 0 \rangle \\
&=(\prod_{a_{i}}\dfrac{1}{N_{a_{i}}!})\int_{-\infty}^{\infty}\dfrac{d\theta_{1}...d\theta_{n}}{(2\pi)^n} \vert F^{\Phi}_{a_{1},...,a_{n}}(\theta_{1},...,\theta_{n}) \vert^{2} e^{iP_{n}x} e^{-iE_{n}t}.
\label{s12}
\ea
\ee
At zero momentum transfer $q=0$, the DSF becomes
\be
\ba
S^{\Phi\Phi}(\omega,q=0)=(\prod_{a_{i}}\dfrac{1}{N_{a_{i}}!})\dfrac{1}{(2\pi)^{n-2}}  \int_{-\infty}^{\infty} d\theta_{1},...,d\theta_{n}\vert F^{\Phi}_{a_{1},...,a_{n}}(\theta_{1},...,\theta_{n}) \vert^{2}\delta(\omega-E_{n})\delta(P_{n})\,,
\label{s13}
\ea
\ee
wsthere $E_{n}=\sum_{i=1}^{n}m_{i}\cosh(\theta_{i}),~P_{n}=\sum_{i=1}^{n}m_{i}\sinh(\theta_{i})$.

\section{The Derivation of The Expressions to Calculate DSF from Different Channels}
\subsection{One Particle Channel}
By Eq.~(\ref{s13}), setting $n=1$, we obtain the DSF for a single particle channel,
\be
S^{\Phi\Phi}_{1}(\omega,q=0)=2\pi\sum_{a_{i}=1}^{8} \vert F^{\Phi}_{a_{i}} \vert^{2}\delta(\omega-m_{a_{i}}).
\label{s14}
\ee
Eq.~(\ref{s14}) shows that the single-particle resonant peaks arise at $\omega=m_{a_{i}}$.

\subsection{Two Particle Channel}
From Eq.~(\ref{s13}), setting $n=2$, we obtain the expression for two particle channel's DSF,
\be
S^{\Phi\Phi}_{2}(\omega,q=0)=\sum_{i\leq j}(\dfrac{1}{2})^{\delta_{a_{i}a_{j}}}\dfrac{\vert F^{\Phi}_{a_{i}a_{j}}(\theta_{1}-\theta_{2}) \vert^{2}}{m_{a_{i}}m_{a_{j}}\vert \sinh(\theta_{1}-\theta_{2}) \vert},
\label{s15}
\ee
where the denominator come from Jacobian, i.e., density of states at zero momentum. We do the variable transformation by defining
\be
\ba
y&=\w -E_{n}=\w-(m_{a_{i}}\cosh\theta_{1}+m_{a_{j}}\cosh\theta_{2}),\\
z&=P_{n}=m_{a_{i}}\sinh\theta_{1}+m_{a_{j}}\sinh\theta_{2}.
\label{s16}
\ea
\ee
Then we get the Jacobian that
\be
d\theta_{1}d\theta_{2}=\dfrac{d y d z}{\left|\begin{array}{cccc}
	\dfrac{\p y}{\p \theta_{1}} &  \dfrac{\p y}{\p \theta_{2}}     \\
	\dfrac{\p z}{\p \theta_{1}} &  \dfrac{\p z}{\p \theta_{2}}   \\
	\end{array}\right| }=\dfrac{d y d z}{m_{a_{i}}m_{a_{j}}\vert \sinh(\theta_{1}-\theta_{2}) \vert}.
\label{s17}
\ee
And the energy-momentum conservation constraints give
\be
\ba
&\omega=m_{a_{i}}\cosh\theta_{1}+m_{a_{j}}\cosh\theta_{2} \\
&0=m_{a_{i}}\sinh\theta_{1}+m_{a_{j}}\sinh\theta_{2}.
\label{s18}
\ea
\ee
From Eq.~(\ref{s18}), $\theta_{1}-\theta_{2}$
can be expressed in terms of $\omega$, $m_{a_{i}}$ and $m_{a_{j}}$, i.e.,
\be
\theta_{1}-\theta_{2}=\arccosh\left(\dfrac{\omega^2-m_{a_{i}}^2-m_{a_{j}}^2}{2m_{a_{i}}m_{a_{j}}}\right).
\label{s19}
\ee
Since all rapidities are all real,
Eq.~(\ref{s19}) immediately implies that the
threshold for the two-particle DSF is
$\omega=m_{a_{i}}+m_{a_{j}}$.

\subsection{Three Particle Channel}
By Eq.~(\ref{s13}), similar to the previous analysis,
three-particle DSF follows by
\be
\ba
S^{\Phi\Phi}_{3}(\omega)=\sum_{i\leq j \leq k}\left(\prod_{a_{i,j,k}}\dfrac{1}{N_{a}!}\right)\dfrac{1}{(2\pi)}\int_{-\infty}^{\infty} d\theta_{k}\dfrac{\vert F^{\Phi}_{a_{i}a_{j}a_{k}}(\theta_{i},\theta_{j},\theta_{k}) \vert^{2}}{m_{a_{i}}m_{a_{j}}\vert \sinh(\theta_{i}-\theta_{j}) \vert},
\label{s20}
\ea
\ee
with the constraints due to energy and momentum conservations
\be
\ba
&\omega=m_{a_{i}}\cosh\theta_{i}+m_{a_{j}}\cosh\theta_{j}+m_{a_{k}}\cosh\theta_{k}, \\
&0=m_{a_{i}}\sinh\theta_{i}+m_{a_{j}}\sinh\theta_{j}+m_{a_{k}}\sinh\theta_{k}.
\label{s21}
\ea
\ee

\subsection{Four Particle Channel}
From Eq.~(\ref{s13}), the expression for the four particle DSF is
\be
\ba
S^{\Phi\Phi}(\omega)=\sum_{i\leq j \leq k \leq l}\left(\prod_{a_{i,j,k,l}}\dfrac{1}{N_{a}!}\right)\dfrac{1}{(2\pi)^2}\int_{-\infty}^{\infty} d\theta_{k}d\theta_{l}\dfrac{\vert F^{\Phi}_{a_{i}a_{j}a_{k}a_{l}}(\theta_{i},\theta_{j},\theta_{k},\theta_{l}) \vert^{2}}{m_{a_{i}}m_{a_{j}}\vert \sinh(\theta_{i}-\theta_{j}) \vert},
\label{s22}
\ea
\ee
with constraints
\be
\ba
&\omega=m_{a_{i}}\cosh\theta_{i}+m_{a_{j}}\cosh\theta_{j}+m_{a_{k}}\cosh\theta_{k}+m_{a_{l}}\cosh\theta_{l} \\
&0=m_{a_{i}}\sinh\theta_{i}+m_{a_{j}}\sinh\theta_{j}+m_{a_{k}}\sinh\theta_{k}+m_{a_{l}}\sinh\theta_{l}.
\label{s23}
\ea
\ee
With $i \leq j \leq k \leq l$ and $\omega$ from 0 to $5m_{1}$,
there are only 3 sets of four particle channels: $m_{1}m_{1}m_{1}m_{1}$, $m_{1}m_{1}m_{1}m_{2}$ and $m_{1}m_{1}m_{1}m_{3}$.
\section{Analysis of The Edge Singularity}
Without loss of generality,
we consider $m_{1}m_{1}$ channel in $D^{xx}$ to show the absence of edge singularity
for two equal-mass channels,
\be
\ba
S_{11}^{xx}(\omega)=\dfrac{1}{2}\left\vert Q^{xx}_{11}(\theta)\times \left(-i\sinh(\dfrac{\theta}{2})\right)\dfrac{G_{\alpha}(2/3,\theta)G_{\alpha}(2/5,\theta)G_{\alpha}(1/15,\theta)}{P_{\alpha}(2/3,\theta)P_{\alpha}(2/5,\theta)P_{\alpha}(1/15,\theta)} \right\vert^{2} \dfrac{1}{m_{1}m_{1}\vert \sinh(\theta) \vert},
\label{s24}
\ea
\ee
where $\theta=\theta_{1}-\theta_{2}$ can be obtained from Eq.~(\ref{s19}).
The product of $Q(\theta)$, $G_{\alpha}(\theta)$ and $P_{\alpha}(\theta)$ is regular,
the singularity given by $1/\sinh(\theta)$
is canceled by the square of $\sinh(\theta/2)$ term
in the form factor, leaving us a regular DSF spectrum.
This can be applied to all two-particle channels with equal masses.

Then we focus on the $m_{a_{i}}m_{a_{j}}(a_{i}\neq a_{j})$ channels with
$|\sinh(\theta/2)|^2$ vanishing in the corresponding DSF.
As such, the edge singularity is given by the $\frac{1}{\vert \sinh(\theta) \vert}$,
\be
\ba
\dfrac{1}{\vert \sinh(\theta_{i}-\theta_{j}) \vert}&=\dfrac{1}{\vert \sinh(\arccosh(\dfrac{\omega^2-m_{a_{i}}^2-m_{a_{j}}^2}{2m_{a_{i}}m_{a_{j}}}))\vert} \\
&=\dfrac{2}{\vert \dfrac{\omega^2-(m_{a_{i}}+m_{a_{j}})^{2}}{m_{a_{i}}m_{a_{j}}\sqrt{\dfrac{\omega^2-(m_{a_{i}}+m_{a_{j}})^{2}}{\omega^2-(m_{a_{i}}-m_{a_{j}})^{2}}}} \vert} \\
\xrightarrow{\omega \sim m_{a_{i}}+m_{a_{j}}} &=\vert \dfrac{1}{\sqrt{(\omega-(m_{a_{i}}+m_{a_{j}}))(\omega+m_{a_{i}}+m_{a_{j}})m_{a_{i}}m_{a_{j}}}}  \vert \\
&\sim \dfrac{1}{\sqrt{\omega-\omega_{0}}}.
\label{s25}
\ea
\ee
where $\omega_{0}=m_{a_{i}}+m_{a_{j}}$, then we have that
\be
\log(S(\omega)) \sim  -\frac{1}{2}(\ln(\omega-\omega_{0})).
\label{s26}
\ee
The prefactor depends on the two-particle state and varies for different DSF expressions.
In the main text, $D_{12}^{xx}$ is shown as an example.

\section{The DSF of $D^{yy}$ and $D^{zz}$}

\label{app:DSFofYYZZ}

\begin{figure*}
	\centering
	\includegraphics[width=11cm]{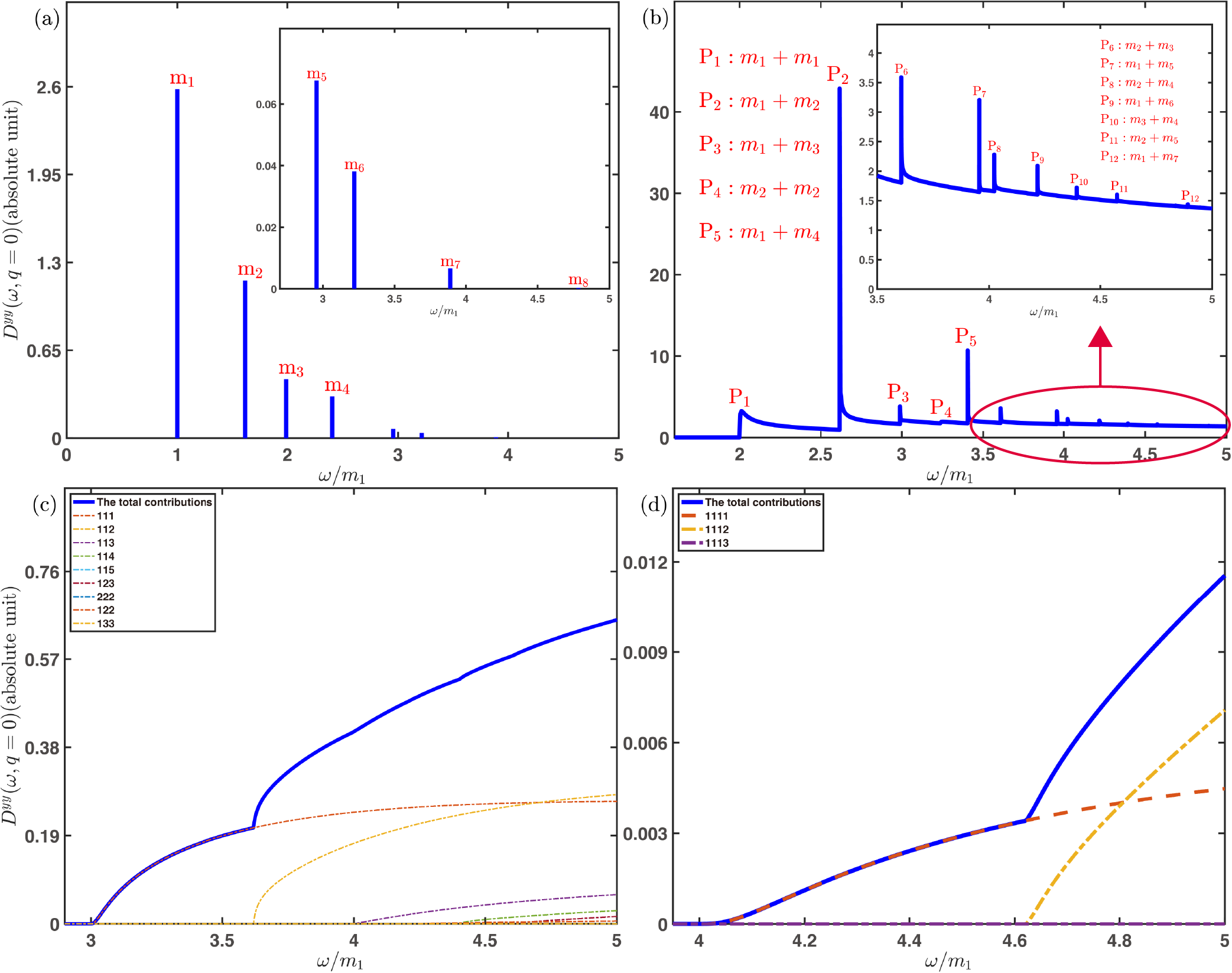}
	\centering
	\caption{The analytical spectra for different particle channels' contributions to the $D^{yy}$. Panel (a)-(d) shows the contribution from single-particle channels to four-particle channels, respectively.}
	\label{Fig.6}
\end{figure*}

\begin{figure*}
	\centering
	\includegraphics[width=11cm]{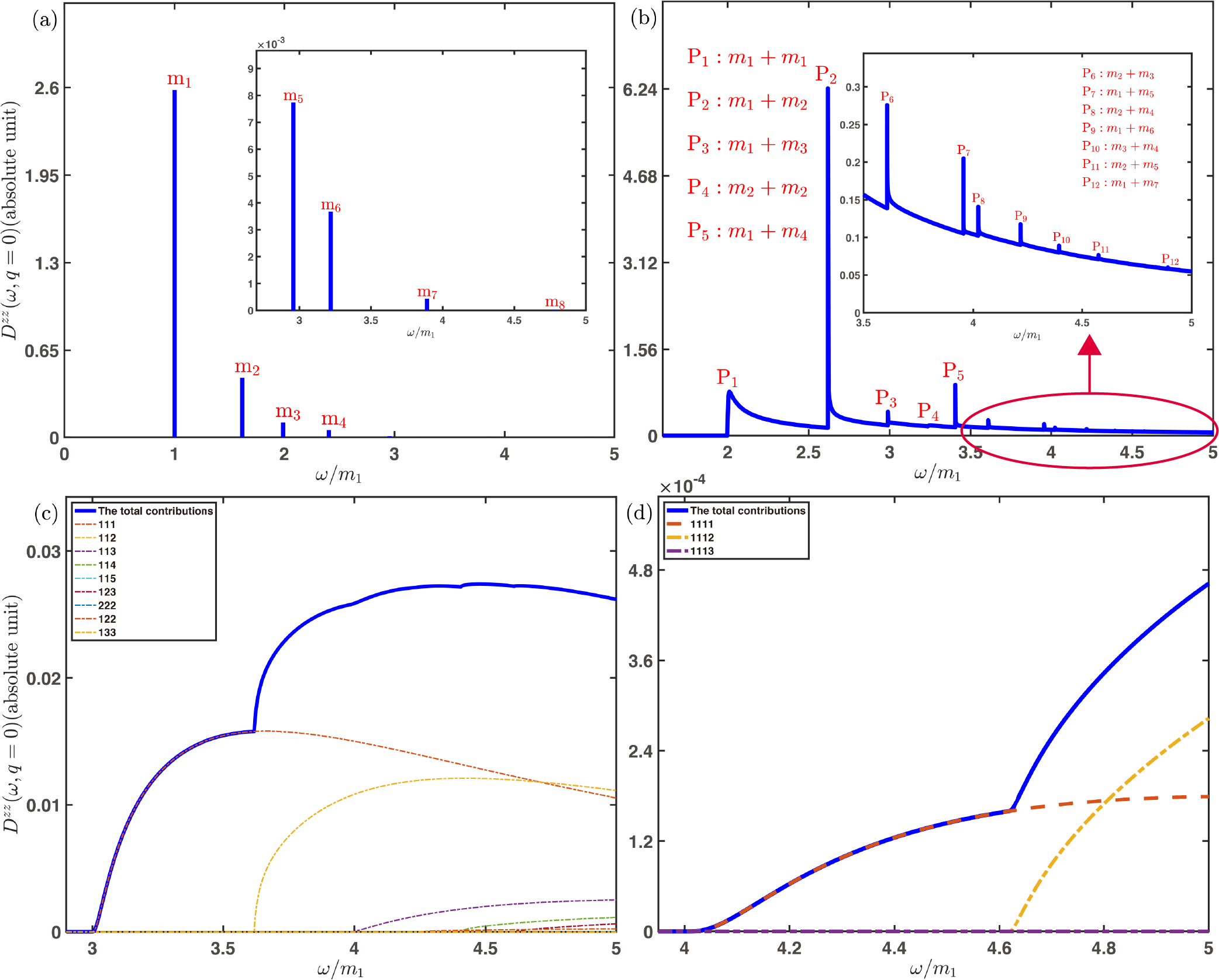}
	\centering
	\caption{The analytical spectra for different particle channels' contributions to the $D^{zz}$. Panel (a)-(d) shows the contribution from single-particle channels to four-particle channels, respectively.}
	\label{Fig.7}
\end{figure*}

\begin{figure*}
	\centering
	\includegraphics[width=11cm]{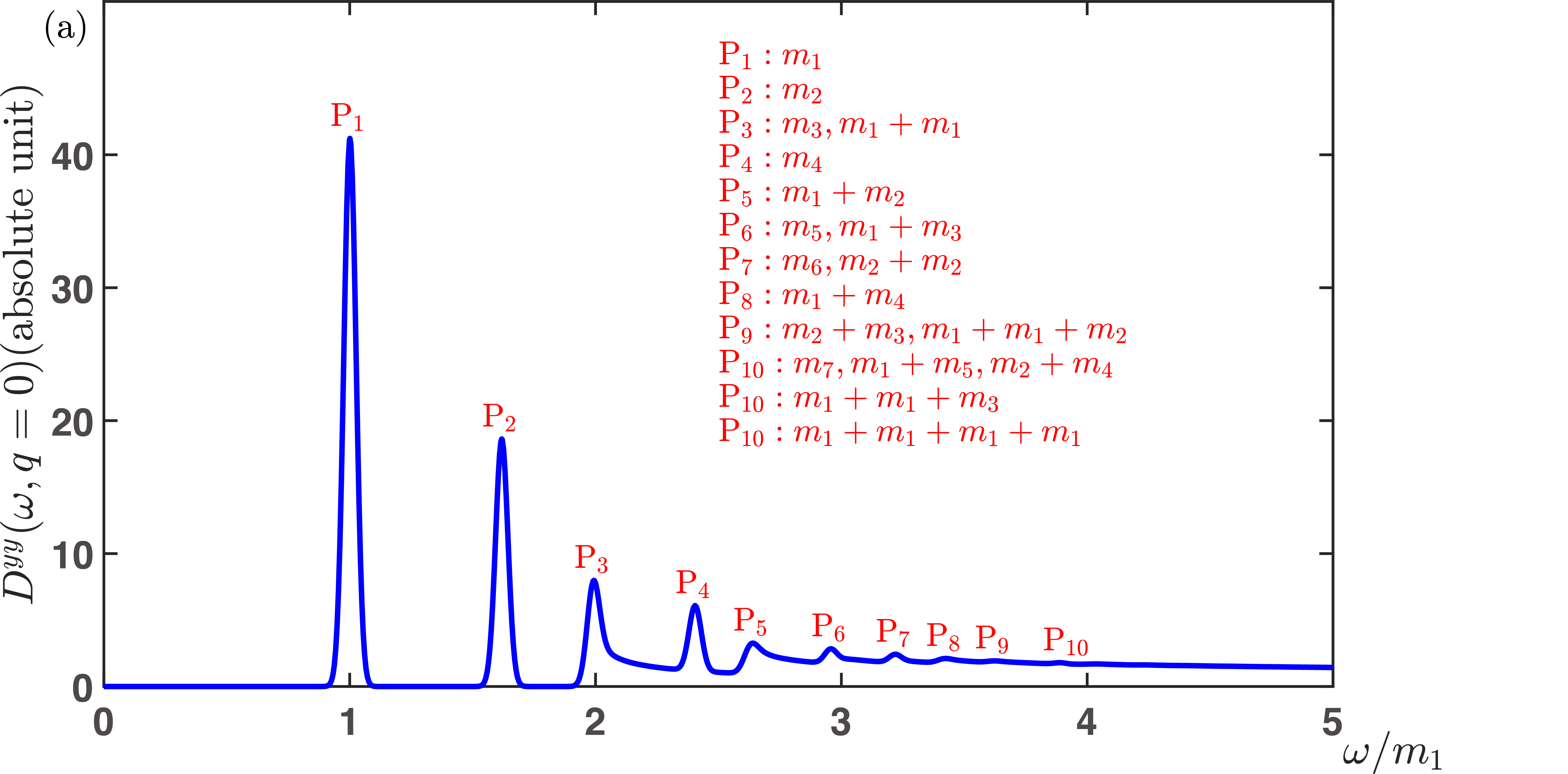}
	\includegraphics[width=11cm]{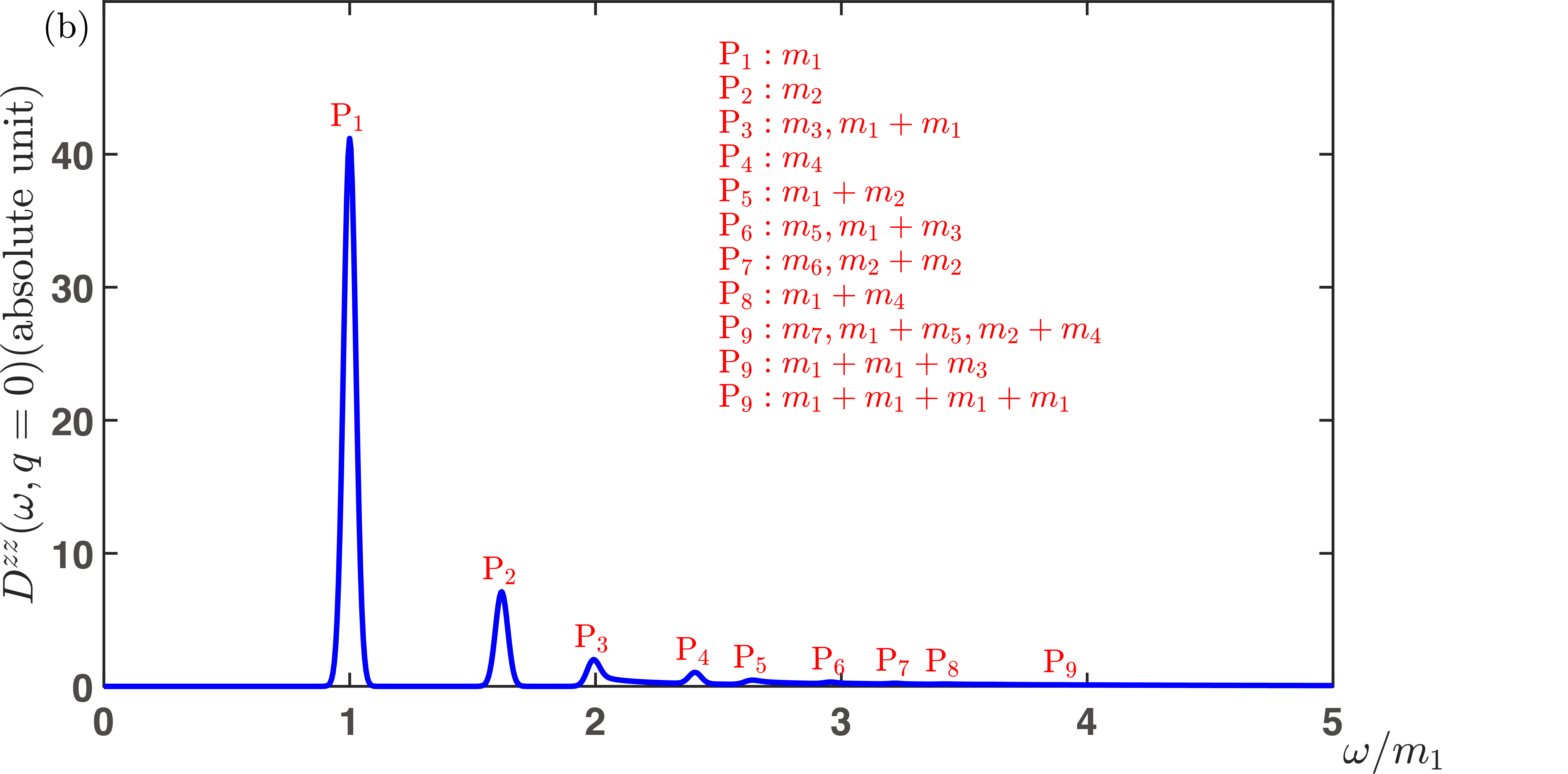}
	\centering
	\caption{The total DSF of (a) $D^{yy}$ and (b) $D^{zz}$ with $0.05m_1$ broadening of analytical data.}
	\label{Fig.8}
\end{figure*}

We also calculated the DSF $D^{zz}$ for the system and $D^{yy}$ can be obtained through the relation $D^{yy}(\w)=\w^{2}D^{zz}(\w)/(4J^{2})$~[\onlinecite{PhysRevLett.113.247201}].
The contributions from different particle channels with energy up to $5m_1$ are shown in Fig.~\ref{Fig.6} and Fig.~\ref{Fig.7} for $D^{yy}$ and $D^{zz}$, respectively.
After broadening the analytical data in Figs.~(\ref{Fig.6},\ref{Fig.7})
energy resolution of $0.05m_1$,
all channels' contributions are combined together and displayed in Fig.~\ref{Fig.8},
which are consistent with the $D^{xx}$ discussed in the main text. 

\end{document}